# Modeling the Fatigue Behavior of Amorphous Polymers

Valeriy V. Ginzburg[1,*], Oleg Gendelman[2], and Alessio Zaccone[3]

[1]Department of Chemical Engineering and Materials Science, Michigan State University, East Lansing, Michigan, USA 48824

[2]Faculty of Mechanical Engineering, Technion, Haifa 32000003, Israel

[3]University of Milan, Department of Physics, via Celoria 16, 20133 Milano, Italy

*Corresponding author: ginzbur7@msu.edu

## Abstract

Prediction of material durability is both very important and very difficult. In many cases, material durability is measured by subjecting a sample to repeating oscillatory cycles (in shear or tension-compression) until it fails in either ductile or brittle fashion. Typically, the stress amplitude is denoted S, and the number of cycles N, so the resulting dependence is known as the SN-curve. For many materials, SN curve has been shown to obey the empirical Basquin's law, $N = AS^{-m}$, where the prefactor $A$ was a function of the temperature, load frequency, and sample history, and the power law $m$ was a real number, usually between 3 and 12. Here, we derive the Basquin's law using a linearized version of the Long's plasticity model and demonstrate that within this framework, $m$ = 3. We also derive the expression for the prefactor $A$. Finally, we show that our theory successfully describes experimental data for three amorphous polymers, PS, PMMA, and PVC.

*Introduction.* All materials (metals, ceramics, or plastics) experience fatigue, i.e., slow degradation of their properties and performance over time, especially if subjected to small yet persistent external forces. One way of characterizing material fatigue is by subjecting a sample to oscillatory shear (or tensile-compressive) load with an amplitude S and measure the number of cycles, N, that elapse before the material fails (in a brittle or ductile fashion).[1–6] Often, the relationship between S and N (the so-called "SN curve") is described by the empirical Basquin's law,[7]

$$N = AS^{-m} \tag{1}$$

Here, $A$ is a phenomenological parameter that depends on the temperature, pressure, load frequency, and sample history, while the power exponent $m$ generally ranging between 3 and 12.[8–12] While there are many theories aiming to predict this behavior from the first principles,[8,9,13] there is still no agreement on the right mechanism of fatigue for glassy polymers. Certainly, predicting fatigue requires detailed understanding of the failure mechanisms (creep vs. crazing etc.)

In this Letter, we propose a simple mean-field analysis that predicts a "universal" Basquin's law for the high-cycle test with R = -1 (i.e., perfect sinusoidal load). In our analysis, the exponent $m$ is predicted to be equal to 3, consistent with the lower-bound limit of experimental data. Furthermore, we provide an estimate for the prefactor $A$. The theory is shown to qualitatively agree with experiments for three polymeric materials: poly(methylmethacrylate) (PMMA), polystyrene (PS), and poly(vinylchloride) (PVC).

*The Model.* Let the material be described using the Maxwell element model. The external load is a sinusoidal function,

$$\sigma(t) = \sigma_0 \, sin(\omega t) \tag{2}$$

The total strain is the sum of the elastic and the viscoplastic components. We are mainly interested in the viscoplastic component, which is given by,

$$E\tau \frac{d\varepsilon_{VP}}{dt} = \sigma(t) \quad (3a)$$

Or

$$\frac{d\varepsilon_{VP}}{dt} = \frac{\sigma_0}{E\tau} sin(\omega t) \quad (3b)$$

After integrating eq 3b, we obtain,

$$\varepsilon_{VP} = -\frac{\sigma_0}{E\omega\tau} cos(\omega t) + C_0 \quad (4)$$

The integration constant $C_0$ should be zero due to symmetry reasons.

Let us consider what happens over the course of one oscillation period ($t = T = 2\pi/\omega$), see Figure 1. The external load is a sinusoidal function (eq 2), as shown in Figure 1a.

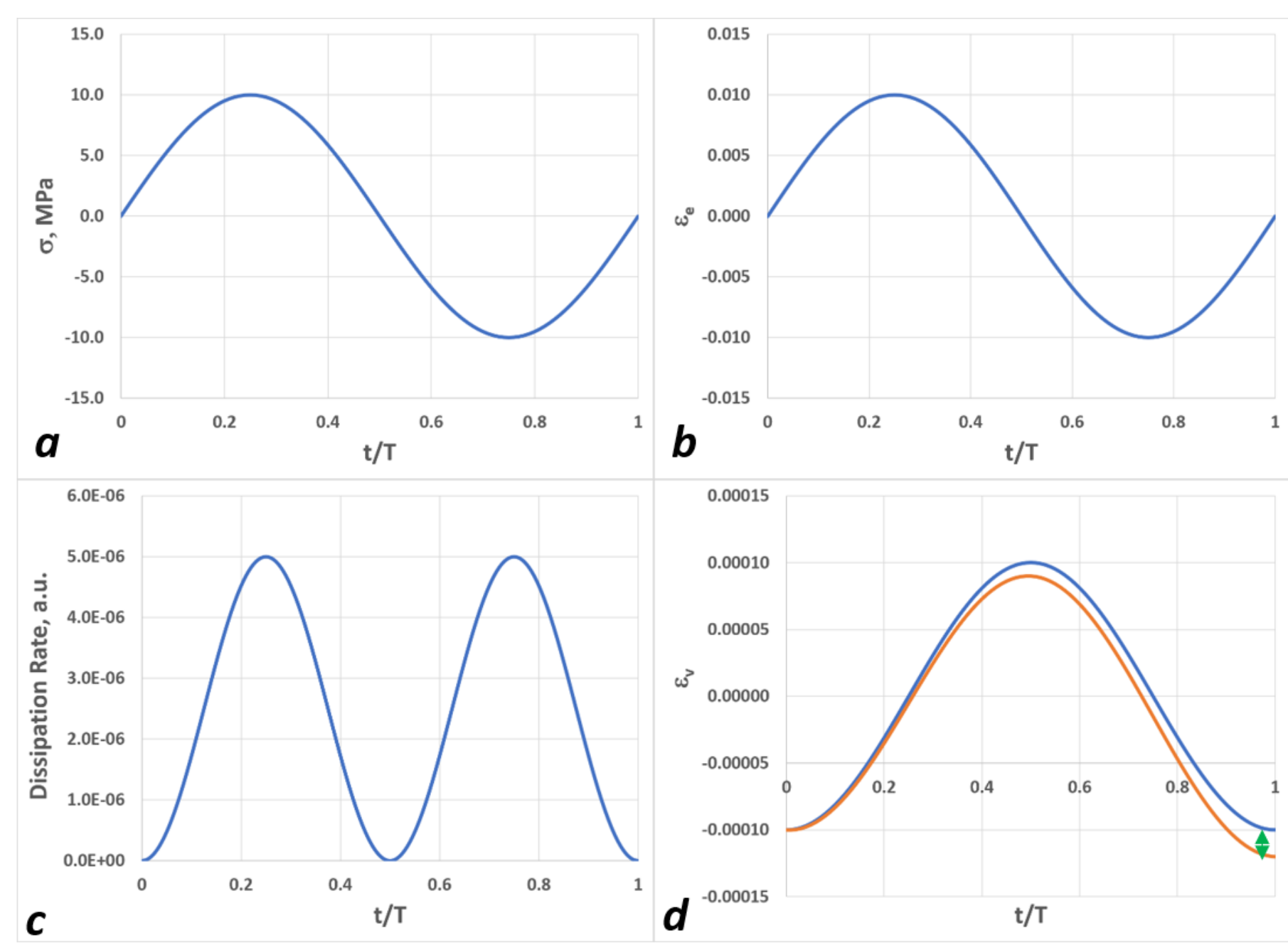

*Figure 1. (a) External stress, (b) elastic strain, (c) dissipation rate, and (d) viscoelastic strain as function of time over one cycle.*

The elastic strain (Figure 1b) is equal to zero at the beginning and the end of the cycle, so there is no accumulated damage. The viscous strain is phase-shifted by $\pi/2$ relative to the elastic strain, but of course the change in the viscous strain between the beginning and the end of the cycle (the blue curve in Figure 1d) is also zero.

However, note that over the course of the cycle, the material absorbs energy (see Figure 1c). Over the course of the first half-period, the material absorbed heat $\Delta Q$, given by,

$$\Delta Q = \int_0^{\frac{\pi}{\omega}} P dt = \int_0^{\frac{\pi}{\omega}} E\tau \frac{\left(\dot{\varepsilon}_{VP}\right)^2}{2} dt = \int_0^{\frac{\pi}{\omega}} E\tau \frac{\left(\frac{\sigma_0}{E\tau}[sin(\omega t)]\right)^2}{2} dt = \frac{\pi}{4\omega}\frac{\sigma_0^2}{E\tau}$$

(5)

(Here, $\Delta Q$ is in units of energy per unit volume).

The absorbed heat increases the temperature of the material in an obvious way,

$$\Delta T = \frac{\Delta Q}{C} \tag{6}$$

Here, *C* is the constant-pressure heat capacity per unit volume. Assuming that the temperature over the course of the first half-period was $T_1$, the temperature over the course of the second half-period was then $T_2 = T_1 + \Delta T$. This leads to the imbalance between the first and the second halves of the cycle (see the orange curve in Figure 1d). The difference between the viscoplastic strain at the beginning of the cycle and the end of the cycle (the green line with arrows in Figure 1d) can be approximated with,

$$\varepsilon_{VP} = \frac{2\sigma_0}{E\omega\tau_1} - \frac{2\sigma_0}{E\omega\tau_2} \tag{7}$$

Here,

$$\tau_1 = \tau_\alpha(T_1) \tag{8a}$$

$$\tau_2 = \tau_\alpha(T_2) = \tau_\alpha(T_1)a_{21} \quad (8b)$$

$$a_{21} = exp\left[\frac{E_a}{RT_2} - \frac{E_a}{RT_1}\right] \approx exp\left(-\frac{E_a(\Delta T)}{RT^2}\right) \quad (8c)$$

(Here, we neglect the difference between $T_1$ and $T_2$ when writing the expression in the denominator).

The relaxation time used here is assumed to be the alpha ($\alpha$) or segmental relaxation time of the glassy polymer. The temperature dependence of the $\alpha$-relaxation time is generally non-Arrhenius,[14–17] so the activation energy $E_a$ used above is itself temperature-dependent.[17,18] Furthermore, at temperatures significantly below $T_g$, $E_a$ is non-equilibrium and depends on the thermomechanical history of the sample and might even depend weakly on the vibration frequency. For the sake of simplicity, we will assume that $E_a$ corresponding to the fatigue experiments is roughly equal to the effective $E_a$ near the glass transition (see below for its estimate).

Substituting eqs 8a – 8c into eq 7, we finally obtain,

$$\varepsilon_{VP,1} = \frac{2\sigma_0}{E\omega\tau_1}\left[1 - \frac{1}{a_{21}}\right] \approx -\frac{2\sigma_0}{E\omega\tau_\alpha(T)}\frac{E_a(\Delta T)}{RT^2}$$

$$= -\frac{2\sigma_0}{E[\omega\tau_\alpha(T)]}\frac{E_a}{RT^2}\frac{\pi}{4}\frac{\sigma_0^2}{EC[\omega\tau_\alpha(T)]}$$

$$= -\frac{\pi}{2}[\omega\tau_\alpha(T)]^{-2}\frac{\sigma_0^3}{E^2CT}\frac{E_a}{RT}$$

(9a)

The above analysis is based on the Maxwell element model, in which the loss tangent is given by, $tan(\delta) = (\omega\tau_\alpha)^{-1}$. Thus, substituting $tan(\delta)$for $(\omega\tau_\alpha)^{-1}$, we can generalize eq 9a to more realistic cases as,

$$\varepsilon_{VP,1} = -\frac{\pi}{2}[tan(\delta)]^2 \frac{\sigma_0^3}{E^2 CT}\frac{E_a}{RT} \quad (9b)$$

The subscript "1" for the strain is to indicate that it is the accumulated strain ("damage") after one cycle. The minus sign indicates that the overall viscoplastic shear strain after one cycle has the direction opposite to the one in which the initial stress was directed.

To determine the fatigue criterion, we need to determine the number of cycles, *N*, before failure. Let us assume that the failure (either brittle or ductile) happens when the total accumulated viscoplastic strain becomes equal to some pre-determined value, $\varepsilon$*. Then,

$$N\frac{\pi}{2}[tan(\delta)]^2 \frac{\sigma_0^3}{E^2 CT}\frac{E_a}{RT} = \varepsilon^* \quad (10)$$

This implies that the S-N curve has a simple power-law asymptotic for small stress amplitudes (here, we use S instead of $\sigma_0$),

$$NS^3 = const. \quad (11)$$

As discussed above, the *SN* curves are typically described using Basquin's law,

$$NS^m = A \quad (12)$$

where *m* is between 3 and 12.[1–3,8] Obviously, the above analysis is purely linear and thus applies only for small S, well below the yield stress. In our model, *m = 3* and $A = \frac{2}{\pi}\frac{RT}{E_a}[tan(\delta)]^{-2}ECT\sigma^*$ (with units Pa$^3$). In writing eq 12, we defined $\sigma^* = E\varepsilon^*$-- this is the "ultimate strength" of the material. The nature of $\sigma^*$is determined based on whether the current conditions correspond to brittle or ductile failure mode.

Let us now apply the model to describe actual experimental SN curves for amorphous thermoplastics. We use the data for PS, PMMA, and PVC – those are the same polymers analyzed in our earlier work on the brittle-ductile transition.[19]

*Results and Discussion.* In Figures 2—4, experimental data for PMMA, PS, and PVC are plotted against our model calculations. The data for PMMA are taken from Okeke et al.,[20] for PS from Iacopi and White,[21] and for PVC from Joseph and Leevers.[22] Details of the tests (temperature and frequency) and model parameters are summarized in Table 1.

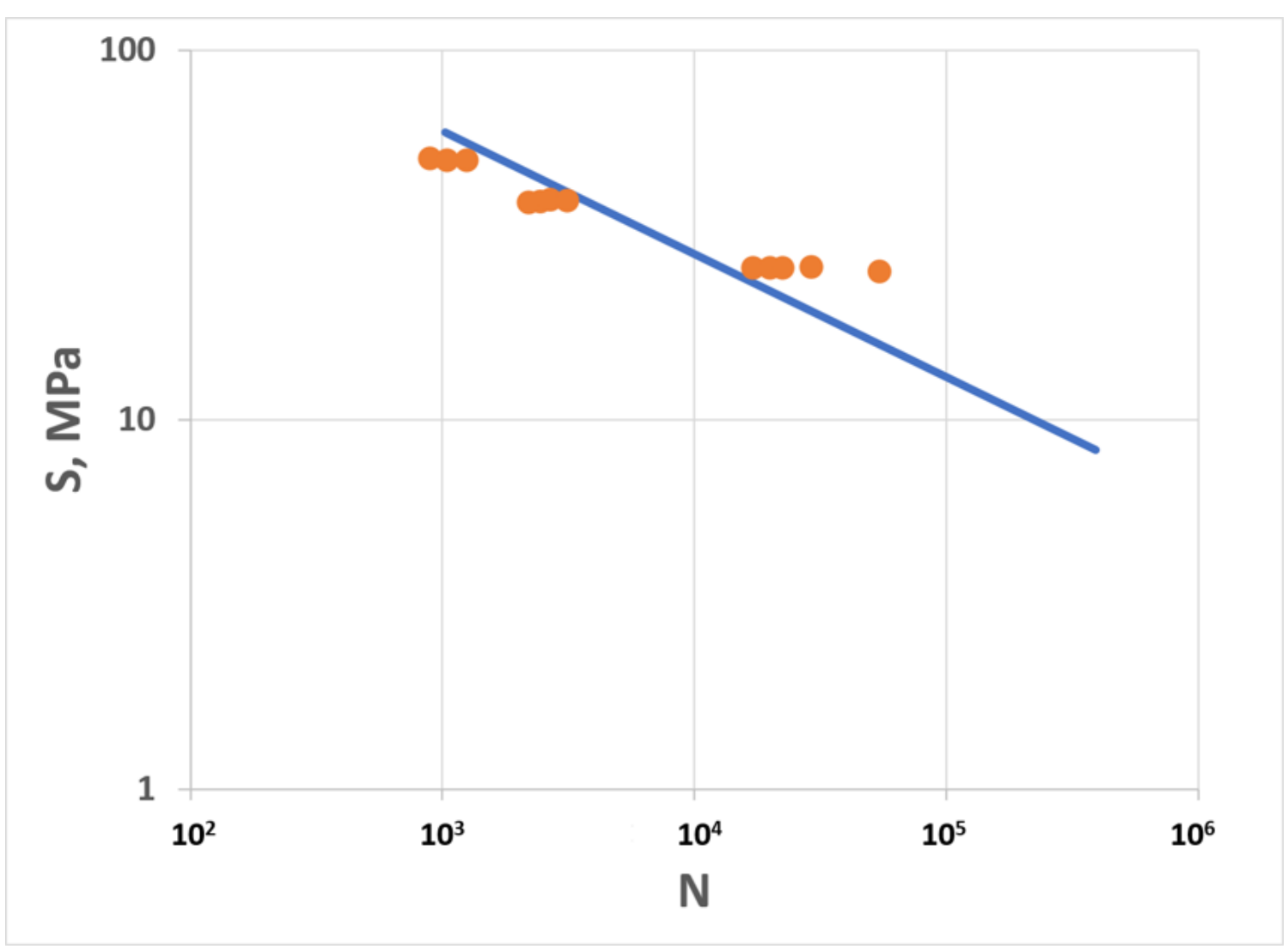

*Figure 2. Experimental (orange circles) and theoretical (blue line) SN curves for PMMA. The data are from Okeke et al.*[20]

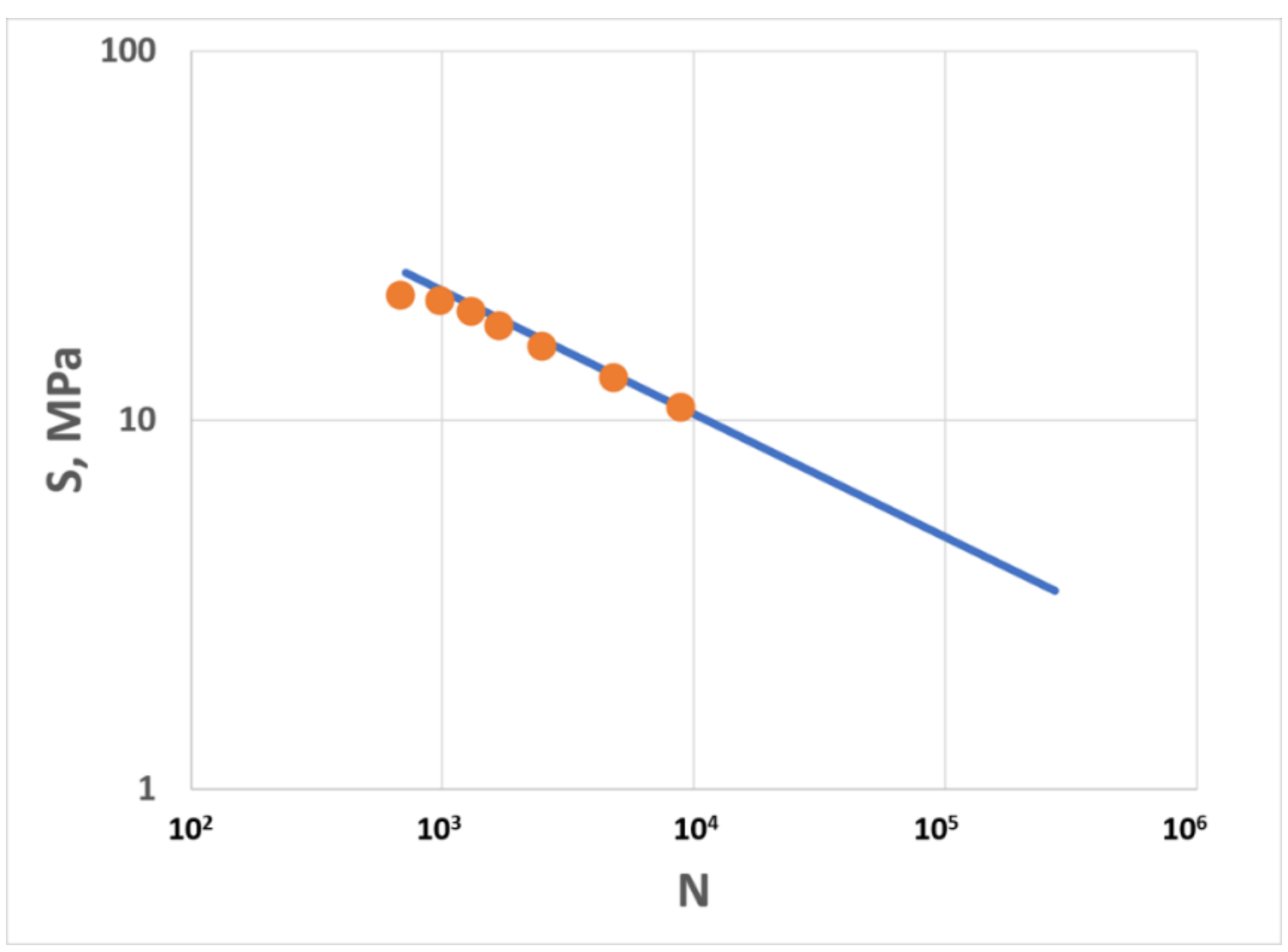

*Figure 3. Experimental (orange circles) and theoretical (blue line) SN curves for PS. The data are from Iacopi and White.[21]*

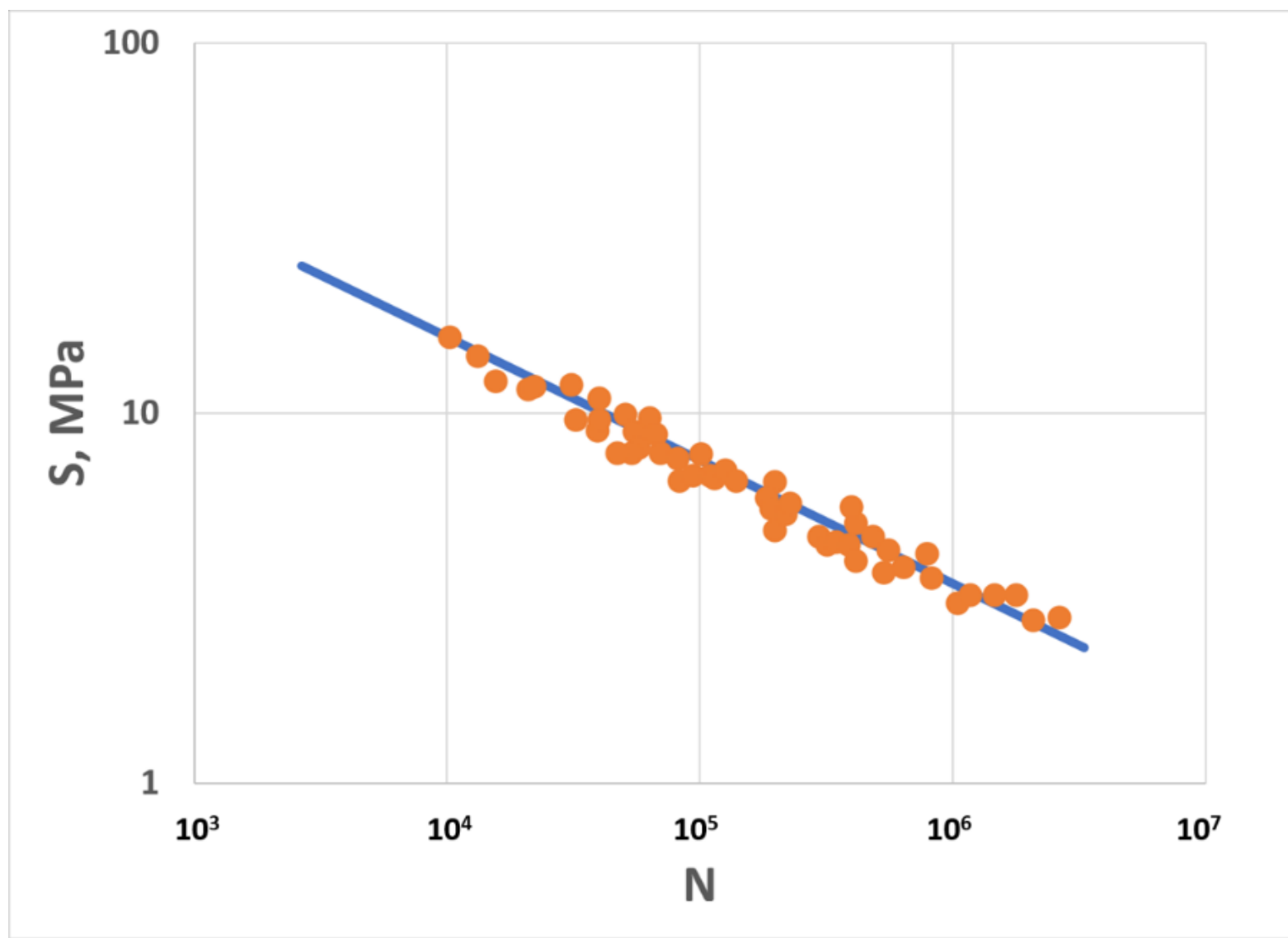

*Figure 4. Experimental (orange circles) and theoretical (blue line) SN curves for PVC. The data are from Joseph and Leevers.[22]*

*Table 1. Experimental conditions, parameters, and calculated Basquin's parameter A*

| Polymer Name | PMMA | PS | PVC |
|---|---|---|---|
| Experimental Conditions | | | |
| T, K | 298 | 298 | 293 |
| $\omega$, rad/s | 282.7 | 2.6 | 6.3 |
| Parameters | | | |
| E, MPa | $3{\cdot}10^3$ | $3{\cdot}10^3$ | $3{\cdot}10^3$ |
| C, J/(K m$^3$) | $1.3{\cdot}10^6$ | $1.3{\cdot}10^6$ | $1.3{\cdot}10^6$ |
| tan($\delta$) | 0.033 | 0.15 | 0.076 |
| $\sigma$*, MPa | 80 | 80 | 80 |
| $E_a$, kJ/mol | 570 | 570 | 570 |
| Calculation Result | | | |
| A, (MPa)$^3$ | $2.2{\cdot}10^8$ | $1.12{\cdot}10^7$ | $4.16{\cdot}10^7$ |

In the above calculations, we assumed the same values of modulus, $E$, heat capacity, $C$, ultimate strength, $\sigma^*$, and activation energy, $E_a$, for all three polymers. To estimate $E_a$, we used the expression $E_a \approx 2.303mRT_g$, where $m$ ~ 100 – 150 is the fragility. Most of these parameters do not vary substantially for the amorphous glassy polymers, although the variations by a factor of two are expected and could strongly impact the final result. The only fitting parameter was the loss tangent, and the best-fit values (between 0.03 and 0.15) are indeed consistent with the experimentally observed loss tangents for glassy polymers.

Overall, given the simplicity of the model, the agreement between theory and experiment is surprisingly good. The predicted power exponent $m = 3$ is reasonably consistent with the observed slope of the SN plots. The prefactor $A$ is a strong function of many variables (temperature, load frequency, modulus, strength, heat capacity, and loss tangent). The loss tangent is a crucial parameter, as it depends significantly on the current conditions (temperature and pressure), as well as the sample history (aging). Thus, it would not be surprising to see a large variability of the prefactor $A$ from sample to sample even for the same polymer chemistry (like PS or PMMA) and the same molecular weight. Furthermore, even within a sample, there can be a distribution of relaxation times – and thus the loss tangent -- if different regions experienced different thermomechanical histories during processing. Such heterogeneity can be one source of variability in the measured N(S). Understanding this variability could be done by employing the methods developed by Weibull,[23,24] where the relaxation time distribution can be determined based on the polymer glass transition theory.

It is important to note that our analysis is based on the idea that the material deforms uniformly. This implies that there are no pre-existing cracks or other defects. If cracks are present, the effective stress amplitude would be increased, at least locally, and the damage growth will have to be re-calculated (though perhaps the same approach could be used with a spatially-dependent S). Similar reservations apply when the material is a polymer-inorganic composite or even a heterogeneous polymer blend.

One artificial assumption in the above model is that the temperature increases abruptly after the first half-cycle and then goes back to the original value (due to the heat conduction) after the second half-cycle. In reality, the temperature is likely increasing monotonically yet very weakly over the course of the fatigue test. Yet, assuming that the heat exchange with the surrounding is sufficiently slow, one can still argue that over the course of each cycle, the second half has a slightly larger temperature than the first half. Thus, the overall approach should be still valid, although the Basquin's constant $A$ would have acquired an additional prefactor.

The current model is the "high-cycle" model, corresponding to the case where the external stress amplitude, S, is significantly smaller than the material strength, $\sigma^*$. As S is increased, we expect that N(S) will begin to deviate downwards from the Basquin's law with $m = 3$, resulting in the smaller apparent slope and the larger apparent $m$. Furthermore, as the damage continues to accumulate, the process is likely to accelerate and deviate from the linear damage accumulation mode assumed here. Nevertheless, this acceleration should not occur until relatively late in the process, so our prediction should still serve as a reasonable first-order estimate. We can improve upon this prediction by employing a nonlinear Maxwell element instead of a linear one,[19] and accounting for the compressed-exponential style accelerated stress relaxations.[25,26] Ultimately, the goal is to relate these parameters to the polymer network morphology and relate the material lifetimes to polymer chemistries. Those are topics for future research.

*Microscopic Interpretation.* The present mean-field model admits a simple microscopic interpretation that may help clarify the physical origin of fatigue damage. Consider a local rearrangement occurring during the first half of the loading cycle. Under the applied stress, a small region of the glass undergoes a nonaffine transformation from an initial configuration $A$ to a new configuration $B$. If the loading and unloading processes are perfectly symmetric, the reverse half-cycle follows exactly the same microscopic pathway, returning the system from $B$ back to $A$. Symbolically, the process can be written as

$$A \rightarrow B \rightarrow A.$$

Although such a rearrangement may be highly nonaffine and dissipative, no permanent structural evolution remains after one complete loading cycle because the original configuration is completely recovered. This situation corresponds to the reversible plastic rearrangements observed experimentally in cyclically sheared amorphous materials.

The essential physical ingredient of the present theory is that viscous dissipation during the first half-cycle generates a small amount of heat. This raises the local temperature by a small amount $\Delta T$, which in turn slightly decreases the segmental ($\alpha$) relaxation time. Consequently, the reverse transformation no longer occurs under exactly the same kinetic conditions as the forward transformation.

Instead of retracing the original microscopic pathway, the unloading stage terminates in a nearby configuration $A'$,

$$A \rightarrow B \rightarrow A',$$

where $A'$ differs slightly from the original configuration $A$. The difference between $A$ and $A'$ represents a finite residual viscoplastic strain accumulated during a single loading cycle. Although this residual strain is extremely small, it is systematically generated during every cycle. The cumulative effect of these infinitesimal irreversible increments naturally leads to fatigue damage and ultimately to material failure.

From this viewpoint, the present model suggests that fatigue originates from a subtle thermally induced breaking of microscopic reversibility. In the absence of dissipation-induced heating, loading and unloading are exactly symmetric and no residual viscoplastic strain is accumulated. Once a small temperature difference develops between the first and second halves of the cycle, however, the relaxation kinetics become slightly different during loading and unloading. This weak asymmetry is sufficient to produce a finite residual strain after every cycle, giving rise to the specific form of the Basquin law derived in the present work.

This interpretation is qualitatively consistent with recent experiments on cyclically sheared amorphous materials, which demonstrated that nonaffine particle rearrangements need not be irreversible. Keim and Arratia[27] showed that a substantial fraction of local rearrangements remains completely reversible over each loading cycle despite dissipating mechanical energy, while irreversible deformation appears only when the loading and unloading trajectories fail to retrace one another exactly. Within the present theory, the dissipation-induced temperature increase provides a natural mechanism for breaking this microscopic reversibility, thereby converting an otherwise reversible rearrangement into one possessing a small irreversible component. Although the present theory is formulated at the mean-field level and does not explicitly resolve individual particle rearrangements, it offers a possible microscopic interpretation of how irreversible damage may emerge from thermally activated asymmetry during cyclic loading.

*Conclusions.* We formulated a simple model that predicts the linear damage accumulation in fatigue tests of glassy polymers. The model successfully recovers the empirical Basquin's law and estimates the power-law exponent $m = 3$; this prediction is consistent with experimental data for several amorphous polymers (PS, PMMA, and PVC). The importance of this model is that it allows one to model the long-term fatigue behavior (millions of cycles) using simple one-time tests (tensile for the modulus and strength, DMA for the loss tangent). It also allows one to immediately evaluate how the number of cycles, N, depends on the test frequency and temperature, thus rapidly accelerating the fatigue testing process. We hope that the model will be of interest to experimentalists and will be tested and validated and/or improved in the near future.